\documentstyle[twocolumn,aps,prl,overcite]{revtex}

\begin{document}
\title{Observational evidence for self-interacting cold dark matter}

\author{David N. Spergel and Paul J. Steinhardt \\
%EndAName
Princeton University, Princeton NJ 08544 USA}

\maketitle

%%%%%%%%%%%%%%%%%%%%%%%%%%%%%%%%%%%%%%%%%%%%%%%%%%%%%%%%%%%%%%%%%%%%%%%%%%
\begin{abstract}
Cosmological models with cold dark matter composed of 
weakly interacting particles  
predict  overly dense cores in 
the centers of galaxies and clusters 
and an overly  large number of
 halos within the Local Group compared to actual observations.
We propose that the conflict can be resolved if the cold
dark matter particles are self-interacting with a large 
scattering cross-section but negligible annihilation or dissipation.
In this scenario,  astronomical observations
may enable us to study dark matter properties that are inaccessible
in the laboratory.  

\end{abstract}
\pacs{PACS number(s): 95.35.+d, 98.35.Gi, 98.62.Ai, 98.80.-k }
%% dark matter, galactic halos, formation of extragalactic
%%objects, cosmology

%%CHANGE
Flat cosmological models with a mixture of ordinary baryonic matter,
cold matter, and cosmological constant (or quintessence) 
and a nearly scale-invariant, adiabatic spectrum of density fluctuations
are consistent with standard inflationary cosmology and
provide an excellent fit to current observations on
large scales ($\gg 1$~Mpc).\cite{Bahcall99}
%%ENDCHANGE
However,  an array of observations 
on galactic and subgalactic scales ($\le$ few  Mpc) 
appears to conflict with 
the structure formation  predicted by analytical calculations
and numerical simulations. The predictions are 
based on the standard view of cold dark
matter as consisting of particles with weak self-interactions, 
as well as weak 
interactions with ordinary matter.

A generic  prediction for weakly self-interacting dark matter,
independent of other details of the 
cosmological model, is that cold dark matter forms
  triaxial  halos 
with dense cores and significant dense 
substructures within the halo. Yet, lensing 
observations of 
clusters\cite{Tyson98} reveal central regions 
(roughly galactic scale) with nearly spherical low density cores.  
%%CHANGE
Dwarf  irregular
%%ENDCHANGE
galaxies appear to have 
low density cores\cite{Moore94,Flores94,DeBlok97,Dalcanton99} 
with much shallower profiles than predicted in numerical 
simulations \cite{Navarro98,Moore98}.  The persistence of bars in high surface brightness galaxies imply that galaxies like our 
own Milky Way also have low density cores \cite{Debattista98}.  Observations of the Local Group reveal less than one 
hundred galaxies \cite{Mateo99}, while numerical 
simulations \cite{Klypin99,Moore99} and analytical 
theory \cite{Press74,Kauffmann93} predict that 
there should be roughly one thousand discrete 
dark matter halos within the Local Group.  

In this paper, 
we propose that the inconsistencies with the standard  
picture may be alleviated if the cold dark matter is 
self-interacting with a large
scattering cross-section but negligible annihilation 
or dissipation.   
The key feature is that the 
mean free path should be in the range 1 kpc to 1 Mpc
%%NEWCHANGE
at the solar radius, where the dark matter density is
about 0.4 GeV/cm$^3$.
%%ENDNEWCHANGE
The large scattering cross-section may be due to  strong, short-range
interactions, similar to neutron-neutron scattering at low-energies, 
or weak interactions mediated by the 
exchange of light particles 
(although not so light as to produce a long-range force). 
%%NEWCHANGE
Depending on the interaction and the mean 
free path, the requisite mass  for the dark matter is 
in the range 1 MeV to 10 GeV.
%%ENDNEWCHANGE
For the purposes of our proposal,
 only two-body scattering effects are important  so
either repulsive or attractive interactions are possible.  
%%ENDCHANGE
Exchanged particles should be massive enough that they are not
radiated by the scattering of dark matter particles
in the halo.

We are led to consider self-interactions because
ordinary astrophysical processes are unlikely to 
 resolve the problems with standard, weakly interacting 
 dark matter.  Consider the dwarf galaxy problem.
One might suppose  that supernova explosions\cite{Dekel86}
could cause the galactic core density to be  made smoother; but, 
while the explosions
 suppress
star formation in dwarf galaxies,  numerical 
simulations\cite{MacLow99} find that starbursts in dwarfs
are very inefficient at removing  gas or matter from the core.
One might also consider whether 
the apparent overabundance of halos found in simulations can 
be explained if the 
low velocity halos form primarily
low surface brightness galaxies\cite{Dalcanton97b}, which 
are difficult to find.  However,
while low brightness galaxy surveys suggest a steeper luminosity function
outside of groups\cite{Dalcanton97a}, even these surveys do not find
enough small galaxies to eliminate the discrepancy between theory and
observations.   If star formation in dwarfs is sufficiently 
suppressed,\cite{Kepner97}
then they should have been detected
%%CHANGE
as  gas clouds in the local group\cite{Blitz99} or external systems.
HI surveys do not find large 
numbers of small isolated gas clouds.\cite{Zwann97}
%%ENDCHANGE
Even if any of the processes were successful in  reducing the number
of visible dwarfs, the dense small halos would still persist.
When these halos fall onto galactic disks, they will heat
the stellar disks and destroy them.\cite{Moore99,Toth92,Weinberg98}  These dense  halos will also settle to the
centers of the central halo and produce a high density core
in galaxies and clusters.  Since the baryon fraction in the
centers of low surface brightness galaxies is low,\cite{Dalcanton97b}
hydrodynamic processes are not likely to alter their
dark matter profiles.\cite{Moore94,Flores94}

The success of the cold dark matter model on large scales suggests
that a modification of the dark matter properties may be the 
best approach for resolving the problems on small scales.
If the dark matter is not cold, but warm (moderately relativistic), 
this alleviates some of these
discrepancies.\cite{Schaeffer88} 
%If the mass of the dark matter is 
%$\sim $ 1 keV, then its initial thermal motion suppresses the formation
%of low mass halos and reduces the central density in more massive halos. 
%Warm dark matter; however, may be too effective in its suppression 
%of small
%scale power. In warm dark matter models, galaxies form primarily
%in clusters and there would not be any Lyman $\alpha$ clouds in low density
%regions.
However, the remarkably good agreement between standard cold dark 
matter (CDM) models and
the observed power spectrum of Lyman $\alpha$ absorbers\cite{Croft99}
likely rules out warm dark matter candidates.

We propose that a better resolution is 
 dark matter that is  cold, non-dissipative,
 but  self-interacting.
There are stringent
constraints on the interactions between dark matter and ordinary
matter\cite{Bern,Stubbs93,Starkman90} and on long range forces between dark 
matter particles\cite{Gradwohl92}. However, 
as long as the dark matter annihilation cross-section is much smaller
than the scattering cross-section, there are relatively few constraints
on short-range dark matter self-interactions.
Carlson, Machacek \& Hall\cite{Carlson92,Machacek94} suggested a
self-interacting dark matter model in which the dark matter
particle is warm rather than cold.  Their model assumed that the 
dark matter plus ordinary matter sum to the critical density predicted
%DAVID: Some shaving here
by inflationary cosmology.  Their purpose was to 
reduce the power on 10 Mpc scales in the dark matter mass spectrum
as required if the  normalization of the spectrum is to agree
fluctuations
measured by the COBE satellite.
 Subsequently, de Laix {\it et al.}\cite{deLaix95}
pointed out that the alteration 
 cannot simultaneously fit the IRAS power
spectrum and the observed properties of galaxies.
Our proposal does not suffer from this problem because
we assume a cosmological model with a low matter density and 
a cosmological constant (or quintessence) which satisfies 
the COBE constraint without self-interactions.
Our proposed self-interactions do not change structure
on the 10 Mpc scale but only on the 1 kpc scale.
Consequently, our model 
satisfies the constraints raised by de Laix {\it et al.}

To be more specific, we suggest that the dark matter particles
should  have a mean free path 
between $\sim$~1 kpc to 1 Mpc 
%%NEWCHANGE
at the solar radius
%%NEWCHANGE
in a typical galaxy
(mean density 0.4 GeV/cm$^3$), for reasons to be explained below.
  For
a particle of mass $m_x$, this implies 
an elastic scattering
cross-section of
\begin{equation}
\sigma_{XX} = 8.1 \times 10^{-25} \, {\rm cm}^{2}  
\left(\frac{m_x}{\rm GeV}\right)
	 \left(\frac{\lambda}{ 1 {\rm Mpc}}\right)^{-1},
\end{equation}
intriguingly similar to that of an ordinary hadron.  (In 
this paper, we consider the case of dark matter particles scattering
only from themselves but, in a forthcoming paper, we consider 
the possibility that dark matter is a stable, neutral hadron.)
% more change 
If the dark matter particles scatter  through 
strong interactions similar to low-energy
neutron-neutron scattering, then
the cross-section is $\sigma = 4 \pi a^2$, where $a$ is the
scattering length. For neutrons, the scattering length is 
more than  100 times its Compton wavelength. 
Using the estimate
$a \approx  100 f m_x^{-1}$, we obtain
\begin{equation}
m_x =  4 \left(\frac{\lambda}{ 1 {\rm Mpc}}\right)^{1/3} f^{2/3} \, GeV.
\end{equation}
Alternatively, the self-interaction may be weak but longer-range,
as in the case of 
the exchange
of a light intermediate vector boson of mass $m_y$, in which case
the cross-section is  $\sigma \approx \alpha_y m_x^2/m_y^4$.
The mass of  the vector boson must be large enough that
there is no dissipation when dark matter particles scatter;
this requires that $m_y >  450 \, {\rm eV} \,  (m_x/GeV)(v/200 km/s)^2$, 
where $v$ is the typical velocity of dark matter particles in the halo.
%%CHANGE
This mass scale for $m_y$ corresponds to a  force that is 
short-range compared
to the dark matter interparticle spacing (about 1 cm in the halo). 
Hence, we need only consider two-body
interactions in our analysis.
%%ENDCHANGE
If $m_y = g m_x$ and $\alpha_y =O(1)$, then the maximum dark
matter mass is
\begin{equation}
m_x < 80 \left(\frac{\lambda}{ 1 {\rm Mpc}}\right)^{1/3} g^{-4/3} \, MeV.
\end{equation}
Beyond what is expressed in the relations above,
there is no significant constraint on how light
the dark matter particles can be. 

The strong self-interaction might occur if the dark matter
consists of particles with a conserved global 
charge (such as a hidden baryon number) interacting through a 
hidden gauge group ({\it e.g.}, hidden color).
  If the gauge group is unbroken,
then the particles experience strong interactions which
can be non-dissipative but the particle number is conserved.
M-theory and superstrings, for example,
suggest the possibility that dark matter fields reside on 
domain walls with gauge fields
  separated from ordinary matter by an extra (small)
dimension\cite{Dvali99a,Dvali99b}. Similar scenarios can be 
constructed in purely
four-dimensional supergravity models.
% yetmorechange
Note that, if the sum of the hidden baryon number and the ordinary
sector baryon number is zero, then $\Omega_{x} = (m_x/m_{proton})
\Omega_{b} \simeq 0.19  \, (m_x/4 {\rm GeV})$ (using $\Omega_b h^2 
= 0.02$ and $h=0.65$).
The particles we suggest 
include light versions
of  WIMPZILLAS\cite{Kolb98}  and Q-balls.\cite{Qball}
%If the dark matter was in thermal equilibrium at one time, there is
%a lower bound on $m_x$ of around 100 keV in order that it behave
%as cold dark matter. However, if it is
%There is no lower limit on the particle mass if the

How does the mean free path of the dark matter  particles
 affect astrophysics?
Since interactions only alter the evolution of cold dark
matter when the density inhomogeneities are large,
the cosmic microwave background (CMB) and large-scale
power spectrum measurements are not sensitive to the self-interactions.
So long as the dark matter is cold ($T_{x} /m_{x}<\Phi$, where
$\Phi$ is the depth of the gravitational potential),
the dark matter will collapse to form a bound halo
regardless of its collisional properties.
If the dark matter mean path were much longer than $\sim $ 1 Mpc, the
typical dark matter particle would not experience any interactions as it
moves through a halo. In this regime, the usual, triaxial cold dark 
matter halo with dense core forms through
gravitational collapse.  On the other hand,
if the dark matter mean free path is much smaller
than 1 kpc, then the dark matter behaves as a collisional gas and this
alters the halo evolution significantly. 
The dark matter will shock: this will heat up
the low entropy material that would usually collapse to form a core and
produce a shallower density profile. Since collisions tend
to make the dark
matter velocity distribution isotropic, 
the halo can not be triaxial and will only be
elliptical if flattened by significant rotation. Since dark halos form
with little angular momentum, if the dark matter is 
not dissipative, then
all halos will be nearly spherical. X-ray observations 
of clusters\cite{Mohr95} reveal
that most halos are moderately ellipsoidal. This implies that the
collision time scale for dark matter near the half-mass
radius of clusters must be longer than the
Hubble time: one of the strongest constraints on this model.
Studies suggest that polar
ring galaxies\cite{Sackett99} are  only mildly triaxial
and oblate with the equatorial plane of the dark halo nearly
coinciding with that of the stellar body.   If
the dark matter has an isotropic distribution function and
the baryons form a disk, then the dark matter will form a slightly
flattened halo.  

In our scenario, we consider a mean free path in the 
intermediate regime,
 larger than 1 kpc, but smaller than $\sim$ 1  Mpc.
Particles in this range have $1 - 10^3$ interactions per Hubble
time in the local neighborhood, which is overdense by $10^6$ relative
to the mean density of the universe.  At the virial radius
of a typical galactic halo, which is overdense by $\sim 200$ relative
to the mean density of the universe, the typical particle
has less than 1 collision per Hubble time.  Thus, near the virial
radius,  halos can
have anisotropic velocity ellipsoids and will be triaxial. However,
in the inner halo of
galaxies, dark matter is collisional. Infalling dark matter is scattered
before reaching the center of the galaxy so that the orbit distribution is
isotropic rather than radial. These collisions increase the entropy of the
dark matter phase space distribution and lead to a dark matter halo profile
with a shallower density profile. 
The characteristic scale for the core would correspond to an ``optical
depth'' of 1, the ``photosphere'' of the dark matter. 

When a dwarf halo with a low velocity dispersion falls into
a larger,  high velocity dispersion halo, 
%%CHANGE
the high velocity particles will scatter off the
low velocity particles.  After the collision, neither particle is
likely to be bound to the dwarf.
%%ENDCHANGE
As dark matter is
slowly removed, the dwarf halo expands, this also makes the halo more
vulnerable to tidal stripping and shock heating. This process 
will slowly evaporate all substructure in the larger halo,
 particularly, in the centers of galaxies, groups
and clusters.  However, near the half-mass radius, the collision
time is longer than the Hubble time, so substructure will only
be destroyed in the inner portions of halos.
 This dark matter evaporation will protect
galaxy disks from dynamical heating 
by collisions with dwarfs in the halo. 
Dwarfs with high
central densities evaporate more slowly as particles at optical depth
greater than $\sim$ 1 are shielded by other particles from collisions.
Intriguingly, all of our Galaxy's dwarf companions have very high phase
space densities.\cite{Mateo99}
As the central density and particle velocities increase,
 not all collisions will lead to particles
being deflected out of the dwarf halo. 
%%CHANGE
%%Some of the particles 
%% DAVID, I ADDED A CONSERVATIVE 
%% ESTIMATE OF THE SUPPRESSION FACTOR. IF THERE IS FURTHER SUPPRESSION,
%% THAT IS OK, BUT THIS IS GOOD ENOUGH FOR OUR PURPOSE. OK?
A small fraction of the particles 
will 
be rescattered within the spheroidal core of the dwarf
and a fraction of the momentum absorbed.  
(The phase space  for capture is suppressed
 by $(\sigma/v)^3$, where
$\sigma$ is the velocity dispersion in the dwarf core and $v$ is the typical
particle velocity in the halo.)
%%ENDCHANGE
This could produce
a ram pressure drag  that can slow the dwarf halo and cause it to 
spiral into the cores of larger halos.
 Numerical studies of this process are required 
to determine whether the observed fraction of denser dwarfs are likely
to survive after a Hubble time.

The halos of large (e.g. $L_*$)  galaxies moving through groups and clusters
are less prone to destruction.  When a cluster dark matter particle
strikes a galaxy dark matter particle, the probability that the recoiling
particles will escape from the galaxy is significantly less than unity.
For cross-sections near the smaller end of our suggested range,
most collisions take place within the half mass radius,
where the escape velocity from a galaxy halo is comparable to the characteristic
recoil velocity.  Thus, many collisions will not lead to recoil energies
large enough to escape. At the smaller end of our cross-section
range, the probability of a typical
particle experiencing a collision during a Hubble time
approaches unity  only for galaxies that fall deep into the
cluster core.  Towards the larger end
of our suggested range, the galaxies are opaque to dark matter
and  the collision products are also likely to
experience multiple collisions within a massive galaxy halo.
Because of these multiple collisions, the collision products
are unlikely to escape from the galaxy.
This effect is more important in massive halos as the optical
depth through a galaxy scales as the one-third power of its mass,
Thus, for large galaxies, the background cluster or group will
primarily heat the entire dark halo rather than evaporate
dark matter from it.

The presence of collisions will lead to energy transport within
the dark matter halo, which eventually leads to core collapse.\cite{Good}
%%NEWCHANGE  minor changes in wording in the rest of this paragraph
We can obtain an estimate of the core collapse time from 
Quinlan's\cite{Quinlan} Fokker-Planck simulations
of the evolution of an isolated cluster of interacting particles
with a central density profile of $r^{-1}$ and an
outer profile of $r^{-3}$.  These models have a temperature
inversion in the core and undergo two stages of core collapse.
During the first stage, the inner region expands as heat is transported inwards.
After 0.1 half-mass relaxation times, the inner 1\% of the mass
has moved out in radius by factor 2.
After roughly 3 half-mass relaxation times, the whole system collapses as heat
is transported from the virial radius outwards.  We suspect that
this second stage is delayed in our cosmological context as the infall
of new material is constantly adding heat near the dark halo photosphere.
If we model our Galaxy as starting with a density profile 
corresponding to the 
Navarro-Frenk-White fit to CDM models\cite{Navarro98} with
$V_{200} = 225$ km~s$^{-1}$, $\Omega_0=0.3$,  $H_0 = 65$~km~s$^{-1}$~Mpc$^{-1}$ 
and dimensionless concentration  parameter
$c=8$, then its half mass radius is 220 kpc.  Since
the density at the half mass radius is 300 times  smaller than the local
dark matter density, the particle mean free path at that radius is
between 0.3 - 300 Mpc and at that radius, the particle is in
the weakly interacting regime.  Thus, for our Galaxy, the core
collapse time  (roughly 3
half-mass relaxation times) is between 4.5 - 6000 Gyr,
 or perhaps
significantly longer  if the collapse stage is delayed by the 
infall of new material.   
Hence, for most  of our range
of parameters, the collapse time for our Galaxy  exceeds the
lifetime of the universe, yet there is sufficient number
of interactions to lower the dark matter density in
the inner 5 kpc of our Galaxy.  As the particle mean free path 
approaches the lower bound (0.3 Mpc) or the upper bound (300 Mpc) at
the half-mass radius, 
 our estimates suggest that one or the other condition
is not satisfied, but more accurate methods are needed to determine
the precise range.  [Note added in proof: A. Burkert (astro-ph/0002409)
has constructed an N-body code that simulates self-interactions and 
obtains results consistent with these estimates based on the 
Fokker-Planck approximation.]

%%ENDNEWCHANGE

%%NEWCHANGE
To summarize, our estimated range of  
$\sigma/m$ for the dark matter is
between 0.45-450~cm$^2$/g or, 
equivalently, $8 \times 10^{-(25-22)}$~cm$^2$/GeV.
Numerical calculations are essential for  checking our 
approximations and refining our estimates.
%%ENDCHANGE
Even without  numerical simulations,
 we can already make a number
of predictions for the properties of galaxies in a self-interacting
 dark matter cosmology: (1) the centers of halos
are spherical; (2) dark matter halos will have cores; and
(3)  there are few dwarf galaxies in groups but 
dwarfs  persist in lower density environments;
and, (4) the halos of dwarf galaxies and galaxy halos in clusters
will have radii smaller than the gravitational tidal radius 
(due to collisional stripping).
Intriguingly, current observations appear to be consistent
with all of these predictions.

\section*{Acknowledgements}
%% check -- I think Nature does not allow acknowledgment of grants
%% but we can put it in the preprint
We would like to thank R. Dave, J. Dalcanton,
G. Dvali, J. Goodman, E. Kolb, J. March-Russell,
 J. Miralda-Escude,
J. Ostriker, J. Peebles,
J. Silk, S. Tremaine, M. Turner and N. Turok for discussions.  
We are grateful to the West Anglia Great Northern
Railway whose train delay provided
the opportunity for initial discussions.
This work
was initiated at the Newton Institute for Mathematical Sciences.  
DNS
acknowledges the NASA Theory program and the NASA MAP Satellite program
for support and PJS was supported by the US Department of Energy grant
DE-FG02-91ER40671.

\end{document}